\documentclass[aps,floatfix,prd,showpacs]{revtex4}
\usepackage{graphicx}
\usepackage{dcolumn}
\usepackage{bm}

\usepackage{amsmath,amsthm,amssymb,epsfig,alltt}

\begin{document}

\def\a{\alpha}
\def\b{\beta}
\def\d{{\delta}}
\def\l{\lambda}
\def\e{\epsilon}
\def\p{\partial}
\def\m{\mu}
\def\n{\nu}
\def\t{\tau}
\def\th{\theta}
\def\s{\sigma}
\def\g{\gamma}
\def\G{\Gamma}
\def\o{\omega}
\def\r{\rho}
\def\z{\zeta}
\def\D{\Delta}
\def\half{\frac{1}{2}}
\def\hatt{{\hat t}}
\def\hatx{{\hat x}}
\def\hatp{{\hat p}}
\def\hatX{{\hat X}}
\def\hatY{{\hat Y}}
\def\hatP{{\hat P}}
\def\haty{{\hat y}}
\def\whatX{{\widehat{X}}}
\def\whata{{\widehat{\alpha}}}
\def\whatb{{\widehat{\beta}}}
\def\whatV{{\widehat{V}}}
\def\hatth{{\hat \theta}}
\def\hatta{{\hat \tau}}
\def\hatrh{{\hat \rho}}
\def\hatva{{\hat \varphi}}
\def\barx{{\bar x}}
\def\bary{{\bar y}}
\def\barz{{\bar z}}
\def\baro{{\bar \omega}}
\def\barpsi{{\bar \psi}}
\def\sp{\sigma^\prime}
\def\nn{\nonumber}
\def\cb{{\cal B}}
\def\2pap{2\pi\alpha^\prime}
\def\wideA{\widehat{A}}
\def\wideF{\widehat{F}}
\def\beq{\begin{eqnarray}}
 \def\eeq{\end{eqnarray}}
 \def\4pap{4\pi\a^\prime}
 \def\xp{{x^\prime}}
 \def\sp{{\s^\prime}}
 \def\ap{{\a^\prime}}
 \def\tp{{\t^\prime}}
 \def\zp{{z^\prime}}
 \def\xpp{x^{\prime\prime}}
 \def\xppp{x^{\prime\prime\prime}}
 \def\barxp{{\bar x}^\prime}
 \def\barxpp{{\bar x}^{\prime\prime}}
 \def\barxppp{{\bar x}^{\prime\prime\prime}}
 \def\barchi{{\bar \varphi}}
 \def\baro{{\bar \omega}}
 \def\bpsi{{\bar \psi}}
 \def\barg{{\bar g}}
 \def\barz{{\bar z}}
 \def\bareta{{\bar \eta}}
 \def\ta{{\tilde \a}}
 \def\tb{{\tilde \b}}
 \def\tc{{\tilde c}}
 \def\tz{{\tilde z}}
 \def\tJ{{\tilde J}}
 \def\tpsi{\tilde{\psi}}
 \def\tal{{\tilde \alpha}}
 \def\tbe{{\tilde \beta}}
 \def\tga{{\tilde \gamma}}
 \def\tchi{{\tilde{\varphi}}}
 \def\barth{{\bar \theta}}
 \def\bareta{{\bar \eta}}
 \def\barom{{\bar \omega}}
 \def\bole{{\boldsymbol \epsilon}}
 \def\bolth{{\boldsymbol \theta}}
 \def\bomega{{\boldsymbol \omega}}
 \def\bolmu{{\boldsymbol \mu}}
 \def\bola{{\boldsymbol \alpha}}
 \def\bolb{{\boldsymbol \beta}}
 \def\bolx{{\boldsymbol x}}
 \def\boly{{\boldsymbol y}}
 \def\bolX{{\boldsymbol X}}
 \def\mathN{{\boldsymbol n}}
 \def\bba{{\boldsymbol a}}
 \def\bby{{\boldsymbol y}}
 \def\bbp{{\boldsymbol p}}
 \def\bbphi{{\boldsymbol \phi}}
 \def\bbA{{\boldsymbol A}}
 \def\mathP{{\mathbb P}}
 \def\mathN{{\boldsymbol N}}
 \def\mathN{{\mathbb N}}
 \def\bbP{{\boldsymbol P}}


\title{Canonical Quantization of Massive Symmetric Rank-Two Tensor in String Theory}

\author{Hayun Park}
\affiliation{
Department of Physics, Kangwon National University, Chuncheon 24341
Korea}
\author{Taejin Lee\footnote{Corresponding author}}
\email{taejin@kangwon.ac.kr}
\affiliation{
Department of Physics, Kangwon National University, Chuncheon 24341
Korea}


\begin{abstract}
The canonical quantization of a massive symmetric rank-two tensor in string theory, which contains two 
Stueckelberg fields, was studied. As a preliminary study, we performed a canonical quantization 
of the Proca model to describe a massive vector particle that shares common properties with the massive 
symmetric rank-two tensor model. By performing a canonical analysis of the Lagrangian, which describes the symmetric 
rank-two tensor, obtained by Siegel and Zwiebach (SZ) from string field theory, we deduced that the Lagrangian 
possesses only first class constraints that generate local gauge transformation. 
By explicit calculations, we show that the massive symmetric rank-two tensor theory is gauge invariant 
only in the critical dimension of open bosonic string theory, i.e., $d=26$. 
This emphasizes that the origin of local symmetry is the nilpotency of the Becchi-Rouet-Stora-Tyutin (BRST)
operator, which is valid only 
in the critical dimension. For a particular gauge imposed on the Stueckelberg fields, 
the gauge-invariant Lagrangian of the SZ model reduces to the Fierz-Pauli Lagrangian of 
a massive spin-two particle. Thus, the Fierz-Pauli Lagrangian is a gauge-fixed version 
of the gauge-invariant Lagrangian for a massive symmetric rank-two tensor. By noting that the Fierz-Pauli Lagrangian is not suitable for studying massive spin-two particles with small
masses, we propose the transverse-traceless (TT) gauge to quantize the SZ model as an 
alternative gauge condition. In the TT gauge, the two Stueckelberg fields can be decoupled from the symmetric rank-two tensor and integrated trivially. The massive spin-two particle can be described 
by the SZ model in the TT gauge, where the propagator of the massive spin-two particle has a well-defined massless limit. 

\end{abstract}


\maketitle

\section{Introduction}

Ever since Dirac \cite{Dirac1959} attempted to quantize Einstein's gravity in the framework 
of the canonical Hamiltonian formulation, the problem of quantization of the spin-two particle has remained 
unsolved. When we perform a canonical quantization of Einstein's gravity, a
number of difficulties may be encountered \cite{Feynman1963,WeinbergPL1964,WeinbergPR1964,WeinbergPR1965}: 
Among others, Einstein's gravity is highly non-linear, and it 
is difficult to apply the usual perturbation theory. If we expand Einstein's gravity in terms of perturbative metrics near the flat Minkowski space-time, we obtain an infinite number of 
interaction terms. Second, because Einstein's gravity does not possess dimensionless coupling, the perturbative quantum theory of Einstein's gravity cannot be renormalized in four dimensions. 
String theory is anticipated to provide reasonable measures to address such difficulties associated with the application of perturbation theory to Einstein's gravity. 
In this respect, it is worth noting that the graviton scattering amplitudes of Einstein's gravity \cite{Scherk1974PLB,Scherk1974NPB,Schwarz1982,Sannan1986,TLeeEPJ2018,TLeeGrav2019} arise from the closed string scattering amplitude in the low-energy approximation where $\ap \rightarrow 0$. By replacing Einstein's gravity with closed string theory in the high-energy region, it is expected that the problem of ultraviolet divergence in quantum gravity could be resolved. 

In the present work, we approach the canonical quantization of the spin-two field from a different angle: we discuss the massive symmetric rank-two tensor, which is closely related to 
the linear theory of massive gravity \cite{FierzP1939}, within the framework of open string theory. 
(For a review of massive gravity, see Refs. \cite{Hinter2012,deRham2014}.) 
In open string theory \cite{SiegelZ1986}, the massive spin-two field is generated as a component of the massive symmetric rank-two tensor multiplet, which additionally contains a vector field and a scalar field. 
These additional fields may be considered as Stueckelberg fields \cite{Stueckelberg1938} to ensure local gauge symmetry, as will be shown later. The massless limit of the massive spin-two field corresponds to the high energy limit in open string theory where $\ap \rightarrow \infty$. Thus, canonical quantization of the massive spin-two field is also intrinsically important when exploring the high energy limit of open string theory \cite{Gross87,Gross88}. 

Introducing Stuckelberg fields to massive spin-two field is not new: If we introduce two Stuckelberg fields to the 
Fierz-Pauli model, describing a massive spin-two paticle, we may obtain a  a well-behaved massive spin-two propagator 
which reduces smoothly to the propagator of massless graviton in the zero-mass limit \cite{Hinter2012,Nibbelink2007}. However, there seems to be no unique way of fixing couplings between Stuckelberg fields to massive spin-two field. 
On the other hand, in string theory the massive spin-two field arises as a component of the massive 
symmetric rank-two tensor multiplet, together with two Stueckelberg fields of which couplings are 
uniquely fixed by the BRST symmetry of string theory. Here, we will discuss this specific form of
massive spin-two field theory, which differs from those previously studied in literature. 

The purpose of this study is twofold: The first is to perform a canonical quantization of the massive symmetric rank-two tensor theory established by Siegel and Zwiebach (SZ) using the BRST-invariant open
string field theory. The second is to clarify its relation to the linear theory of massive gravity studied by Fierz and Pauli (FP) and to propose an alternative, yet equivalent model, that may not share the undesired features of the FP model in the massless limit. We note some of the problems of the FP model in the massless limit from the point of view of canonical quantization and propose an alternative theory for the massive spin-two field,
which can be obtained from the massive symmetric rank-two tensor theory, by choosing the transverse-traceless (TT) gauge \cite{Gravitation1973,Brandt2007,Gribosky1989,Rebhan1991,Blasi2017}. The problematic features of the FP model, such as the discontinuity of the massless limit \cite{vanDam1970,Zakharov1970}, which is termed as the van Dam-Veltman-Zakharov (vDVZ) discontinuity, may be attributed to inadequate choice of gauge fixing for the massive symmetric rank-two tensor theory of Siegel and Zwiebach with small mass. By an explicit construction, we show that in the TT gauge, the propagator of the massive spin-two field smoothly reduces to that of a massless graviton in the massless limit without encountering any discontinuities.

\section{PRELIMINARY: Canonical Quantization of the Proca Model}

As a preliminary study, it may be instructive to first discuss the Proca model \cite{Proca1936} that describes a massive vector particle because it is simple and shares many similarities with the massive symmetric rank-two tensor.
A massive spin-one field is described by the Proca model whose Lagrangian is given as follows:
\beq \label{proca1}
\mathcal{L}_1 = -\frac{1}{4}F_{\m \n}F^{\m \n}-\half m^2 A_\m A^\m 
\eeq
where $F_{\m\n} = \p_\m A_\n - \p_\n A_\m$. Throughout this paper, we will use $\eta_{\m\n} = diag(-, +, \dots, +)$ 
as the $d$-dimensional space-time metric. Performing the Hamiltonian analysis of the Proca action, we find two second class constraints \cite{Vytheeswaran1997,Park1997}, namely $\varphi_1$ and 
$\varphi_2$:
\begin{subequations} 
\beq
\varphi_1 &=& \Pi_0 = 0, \\
\varphi_2 &=& \p_i \Pi_i - m^2 A_0 = 0, \\
\left[\varphi_1(\bolx), \varphi_2(\bolx^\prime) \right]_{PB} &=& m^2 \d(\bolx -\bolx^\prime) \not= 0. 
\eeq
\end{subequations} 
where $\Pi_\m$, $\m =0, 1, \dots, d-1$ are canonical momenta that are conjugate to $A^\m$. This pair of 
second class constraints compels us to use the Dirac brackets instead of the Poisson brackets in the Hamiltonian formulation of the model. Usually, canonical quantization with second class constraints 
often turns out to be cumbersome and complicated \cite{Dirac1959}. 
This problem may be resolved by introducing an auxiliary 
scalar field called Stueckelberg field \cite{Stueckelberg1938}. With the help of a Stueckelberg field, $\phi$, we can recast the model into a more convenient form as follows:
\beq \label{proca2}
\mathcal{L}_2 = -\frac{1}{4}F_{\m \n}F^{\m \n}-\half m^2 \left(A_\m + \p_\m \phi \right)^2.
\eeq
This Lagrangian Eq. (\ref{proca2}) is now invariant under a local gauge transformation 
\beq \label{gauge}
\d A^{\m} = \p^\m \e,~~~\d \phi = - \e .
\eeq
By imposing a gauge fixing condition $\phi =0$, the gauge-invariant Lagrangian Eq. (\ref{proca2}) reduces to the Proca 
Lagrangian Eq. (\ref{proca1}). Thus, the Proca Lagrangian Eq. (\ref{proca1}) can be understood as a gauge-fixed version of 
the gauge-invariant Lagrangian Eq. (\ref{proca2}). 

The advantage of using the gauge-invariant Lagrangian is that we do not need to employ the Dirac brackets to perform the canonical analysis: The Hamiltonian system of the gauge-invariant model possesses only first class constraints, which generate a local gauge transformation. Because we can use the usual Poisson brackets to carry out the Hamiltonian formulation, quantization of the system is simpler and easier than that with the Dirac brackets. Defining the canonical momenta as 
\beq
\Pi^\m &=& \frac{\p {\cal L}}{\p \dot A_\m}, ~~~ \pi_\phi = \frac{\p {\cal L}}{\p \dot \phi},
\eeq 
we find 
\beq
\Pi^i = F_{0i}, ~~~ \pi_\phi = m^2 \left(\dot \phi + A_0 \right)
\eeq 
and a primary constraint 
\beq \label{primary}
\varphi_1 = \Pi^0 = 0.
\eeq 
The Hamiltonian density for the system may be written as follows:
\beq \label{Hamil}
{\cal H} &=& \pi_i\dot{A}^i+\pi_{\phi}\dot{\phi} -\mathcal{L} \nn\\
&=& \half \Pi_i\Pi^i+\frac{1}{4}F_{ij}F^{ij}+\frac{1}{2m^2}\pi_{\phi}^2
+\half m^2(A_i+\p_i \phi)^2+A^0(\p_i \Pi^i+\pi_{\phi}).
\eeq 
It is clear from Eq. (\ref{Hamil}) that $A^0$ is a Lagrangian multiplier. 
Taking a commutator between the primary constraint Eq. (\ref{primary}) and 
the Hamiltonian, we obtain a secondary constraint as follows:
\beq
\varphi_2 = \left[\pi_0, H\right]_{PB} = -(\p_i \Pi^i+\pi_{\phi})= 0.
\eeq
The following simple algebraic relations confirm that the closure property is satisfied:
\beq
\left[\varphi_1, H \right] = \varphi_2, ~~~ \left[\varphi_2, H \right] = 0, ~~~
\left[\varphi_1, \varphi_2 \right] = 0. 
\eeq
Using this algebraic constraint, we may construct a gauge generator as follows:
\beq
\Omega_\e (t) &=& \int d^{d-1} \bolx \left(\dot \e \varphi_1 + \e \varphi_2 \right) \nn\\
&=& \int d^{d-1} \bolx \Bigl\{\dot \e \Pi^0 - \e \left(\p_i \Pi^i + \pi_\phi \right)
\Bigr\},
\eeq  
which then generates the local gauge transformation of Eq. (\ref{gauge}) as
\beq
\d A^\m = \left[A^\m, \Omega_\e \right] = \p^\m \e, ~~~ 
\d \phi = \left[ \phi, \Omega_\e \right] = -\e . 
\eeq 

Having performed the canonical analysis, we may choose a suitable gauge condition to fix the degrees of freedom of the local gauge. Various choices of the gauge fixing conditions are available: 
We may choose the usual gauge conditions such as the Lorenz gauge condition 
$\p_\m A^\m = 0$, Coulomb gauge condition ${\boldsymbol \nabla} \cdot {\boldsymbol A} = 0$,
or axial gauge condition $\hat {\boldsymbol n} \cdot {\boldsymbol A} = 0$.
Alternatively, we may choose a gauge fixing condition that can be imposed on the Stueckelberg field as follows:
\beq \label{gauge2}
\chi = \phi = 0.
\eeq 
This gauge condition also fixes the gauge degrees of freedom appropriately and causes the gauge-invariant Lagrangian of Eq. (\ref{proca2}) to reduce to the Proca model, i.e., Eq. (\ref{proca1}). 
As $ \d \chi(\bolx)/ \d \e(\bolx^\prime) = - \d(\bolx - \bolx^\prime)$, this gauge fixing does not produce a nontrivial Faddeev-Popov ghost term. It may thus be appropriate to call this gauge condition (Eq. (\ref{gauge2})) 
as the Proca gauge.  

In the Proca gauge $\phi=0$, the generating function may be written as   
\beq
Z = \int D[A] D[\phi] \exp \left\{i \int d^d x \left(-\frac{1}{4}F_{\m \n}F^{\m \n}-\half m^2 \left(A_\m + \p_\m \phi \right)^2 - \frac{\l}{2} \phi^2 \right)\right\}
\eeq 
where $\l$ is a gauge parameter \cite{Itzykson}. If we are interested in constructing a propagator for a massive spin-one particle using the Stueckelberg field, it would be convenient to treat the vector field and the Stueckelberg field as components of a multiplet $\Psi^t = (A^\m, \phi)$: In terms of the multiplet, the Proca action may be written as
\beq
{\cal A} &=& \half \int \frac{d^d p}{(2\pi)^d} \Psi^t {\cal O} \Psi, \\
{\cal O} &=& \biggl( \begin{array}{cc}
-\eta_{\m\n}(p^2+m^2)+p_\m p_\n & -im^2p_\m \\ im^2 p_\n & -\l -m^2p^2
\end{array} \biggr).
\eeq   
Inverting the operator ${\cal O}$, we obtain the propagator $G$ for $\Psi$ as follows: 
\beq
G = \frac{1}{i}{\cal O}^{-1} = \biggl( \begin{array}{cc}
\frac{i}{p^2+m^2}\left(\eta^{\m \n}+\frac{p^\m p^\n}{m^2}\right)+\frac{i}{\l}p^\m p^\n & \frac{1}{\l}p^\n
\\ -\frac{1}{\l}p^\m & \frac{i}{\l}
\end{array} \biggr).
\eeq
Here, we use the block matrix inversion 
\beq
\biggl( \begin{array}{cc}
A & B \\ C & D
\end{array} \biggr)^{-1}
=\biggl( \begin{array}{cc}
A^{-1}+A^{-1}B(D-CA^{-1}B)^{-1}CA^{-1} & -A^{-1}B(D-CA^{-1}B)^{-1} 
\\ -(D-CA^{-1}B)^{-1}CA^{-1} & (D-CA^{-1}B)^{-1}
\end{array} \biggr).
\eeq
In the limit where $\l \rightarrow \infty$, the Stueckelberg field $\phi$ completely decouples from the gauge fields $A^\m$, and the propagator reduces to 
\beq \label{propagator}
G =
\biggl( \begin{array}{cc}
\frac{i}{p^2+m^2}\left(\eta^{\m \n}+\frac{p^\m p^\n}{m^2}\right) & 0
\\ 0 & 0
\end{array} \biggr) . 
\eeq
Therefore, in the Proca gauge $\phi=0$ and Eq. (\ref{gauge}), we only need to consider the gauge fields $A^\m$, whose propagator is given by 
\beq
G^{\m\n}_A =\frac{i}{p^2+m^2}\left(\eta^{\m \n}+\frac{p^\m p^\n}{m^2}\right). \label{procagauge}
\eeq 

One of the interesting points that we should note is the massless limit of the Proca model. From the Lagrangian of the Proca model Eq. (\ref{proca1}), we expect that in the massless limit, 
$m \rightarrow 0$, and the Proca model reduces to the $U(1)$ gauge theory. However, in quantum theory, this limit may not be considered in a straightforward manner: The propagator in Eq. (\ref{propagator}) is ill-defined in the massless limit. We may encounter similar difficulties when quantizing the Proca action directly, i.e.,
Eq. (\ref{proca1}), which contains the second class constraints. Because the Dirac bracket is defined \cite{Vytheeswaran1997} as
\begin{subequations}
\beq
\left[F(\bolx), G(\bolx^\prime)\right]_{DB} &=& \left[F(\bolx), G(\bolx^\prime)\right]_{PB} - \int d^{d-1} \boly\int d^{d-1} \boly^\prime \left[F(\bolx), \varphi_a(\boly)\right]_{PB}
E^{-1}_{ab}(\boly, \boly^\prime) \left[\varphi_b(\boly^\prime), G(\bolx^\prime)\right]_{PB}, \\
E_{ab}(\bolx, \bolx^\prime) &=& \left[\varphi_a(\bolx), \varphi_b(\bolx^\prime)\right]_{PB} = \left(
\begin{array}{cc} 0 & m^2 \\ -m^2 & 0 \end{array} \right) \d(\bolx-\bolx^\prime), \\
E_{ab}^{-1}(\bolx, \bolx^\prime) &=& \left(
\begin{array}{cc} 0 & -1/m^2 \\ 1/m^2 & 0 \end{array} \right) \d(\bolx-\bolx^\prime),
\eeq     
\end{subequations}
the Dirac bracket is also ill-defined in the massless limit. This difficulty may be attributed to the fact that the numbers of degrees of freedom of the two theories differ from each other: The Proca model 
describing a massive vector particle has $d$ degrees of freedom in $d$ dimensions, whereas the $U(1)$ gauge theory that describes a massless gauge particle possesses only two degrees of freedom.
 
In order to explore the massless limit of the Proca model, it may be more appropriate to choose the usual Lorenz gauge $\p_\m A^\m =0$. In this covariant gauge, the Proca action with the Stueckelberg
field Eq. (\ref{proca2}) may be read as  
\beq \label{modellorenz}
S &=& \int d^d x \left\{
-\frac{1}{4}F_{\m \n}F^{\m \n}-\frac{m^2}{2} A_\m A^\m - \frac{\l}{2} \left(\p_\m A^\m \right)^2 
- \frac{1}{2} \p_\m \phi \p^\m \phi \right\}
\eeq 
where we scale the Stueckelberg field as $\phi \rightarrow \phi/m$. As the vector field $A_\m$ decouples from the Stueckelberg field, we can obtain their propagators separately as follows: 
\begin{subequations}
\beq
G_A^{\m\n} &=& \frac{i}{p^2+m^2}\left(\eta^{\m \n}+ \a \frac{p^\m p^\n}{p^2}\right), ~~~
G_\phi = \frac{i}{p^2} , \\
\a &=& \frac{p^2(1-\l)}{p^2+m^2-p^2(1-\l)} . 
\eeq 
\end{subequations}
In the limit where $\l \rightarrow \infty$, we find that the propagator for $A_\m$ satisfies the Lorenz gauge condition:
\beq \label{promassive}
G_A^{\m\n} &=& \frac{i}{p^2+m^2}\left(\eta^{\m \n} - \frac{p^\m p^\n}{p^2}\right), ~~~ 
G^{\m\n}_A p_\m = G^{\m\n}_A p_\n = 0. 
\eeq 
In the Lorenz gauge, the massive vector particle is described by $A_\m$, and satisfies the gauge condition $\p_\m A^\m =0$ and massless scalar field $\phi$.  
It is worth noting the difference between the propagator for the massless $U(1)$ gauge field in the Lorenz gauge 
\beq
G_{U(1)}^{\m\n} &=& \frac{i}{p^2}\left(\eta^{\m \n} - \frac{p^\m p^\n}{p^2}\right) 
\eeq
and the propagator for the massive vector of Eq. (\ref{promassive}). 
In contrast to the model of Eq. (\ref{proca1}) in the Proca gauge, 
where the propagator for $A^\m$ becomes singular in the massless limit, 
the model in the Lorenz gauge has a well-defined propagator for $A^\m$, i.e., Eq. (\ref{promassive}), which 
reduces to the propagator of the massless $U(1)$ gauge particle in the massless limit.  
Thus, the model in the Lorenz gauge Eq. (\ref{modellorenz}) 
appears to be more suitable than the conventional Proca model of Eq. (\ref{proca1}) for exploring 
the small-mass limit of the massive vector particle. The two models, Eq. (\ref{proca1}) and Eq. (\ref{modellorenz}), are gauge equivalent to each other. If we further integrate the massless scalar Stueckelberg field 
$\phi$, we get the following Lagrangian for a massive vector field that satisfies the Lorenz gauge condition:
\beq
{\cal L} &=& -\frac{1}{4}F_{\m \n}F^{\m \n}-\frac{m^2}{2} A_\m A^\m - \frac{\l}{2} \left(\p_\m A^\m \right)^2.
\eeq

\section{Massive Symmetric Rank-Two Tensor in String Theory}

We are now ready to discuss the canonical quantization of the massive symmetric rank-two tensor. 
In string theory, massive spin-two fields arise in the spectrum of open strings rather than closed strings as a part of massive rank-two tensor multiplet, which may be described by the following Lagrangian \cite{SiegelZ1986}:
\beq \label{LSZ}
\mathcal{L}_{SZ} &=& \frac{1}{4}h_{\m \n}(\p^2-m^2)h^{\m  \n}+\half B_\m (\p^2 -m^2)B^\m -\half \eta (\p^2-m^2)\eta \nn\\
&& +\half (\p^\n h_{\m \n}+\p_\m \eta -mB_\m)^2+\half(\frac{m}{4}h+\frac{3}{2}m\eta+\p_\m B^\m)^2
\eeq
where $h = h^\s{}_\s$.
Here, $B^\m$ and $\eta$ are Stueckelberg fields that ensure the local gauge symmetry of ${\cal L}$.
By algebraic manipulation, we find that this Lagrangian is invariant under the following local gauge transformation:
\beq \label{gaugespin2}
\d h_{\m \n} = \p_\m \e_\n + \p_\n \e_\m -\half m \eta_{\m \n}\e,~~~
\d B_\m = \p_\m \e + m\e_\m,~~~\d \eta = -\p_\m \e^\m+\frac{3}{2}m\e . 
\eeq
It is worth noting that the explicit calculations given in the Appendix show that the gauge invariance works only for the critical dimension $d = 26$. This reminds us that the origin of the gauge symmetry 
is the nilpotency of the Becchi-Rouet-Stora-Tyutin (BRST) charge \cite{BRS,Tyutin} of open bosonic string theory, which requires that $d = d_{\rm critical} = 26$. 

The equations of motion for fields $h_{\m\n}$, $B_\m$, and $\eta$ are respectively read from the Lagrangian as follows:
\begin{subequations}
\beq
\half (\p^2-m^2) h_{\m \n}- \p_{(\m} \p^\l h_{\n) \l}+m\p_{(\m}B_{\n)}-\p_\m \p_\n \eta
+\eta_{\m \n}\left(\frac{m^2}{16}h+\frac{3}{8}m^2\eta+\frac{m}{4}\p_\l B^\l \right) &=& 0, \label{eqm1}\\
\p^2 B_\m-\p_\m \p_\n B^\n-m\p^\n h_{\m \n}-\frac{m}{4}\p_\m h-\frac{5}{2}m\p_\m \eta &=&0, \label{eqm2} \\
-2\p^2 \eta+\frac{13}{4}m^2 \eta-\p_\m \p_\n h^{\m \n}+\frac{3}{8}m^2 h+\frac{5}{2}m\p_\m B^\m &=& 0 \label{eqm3}.
\eeq
\end{subequations}
If we choose a gauge 
\beq \label{FPgauge}
B^\m = 0, ~~ \eta = - \half h^\m{}_\m, 
\eeq 
to fix the gauge degrees of freedom of Eq. (\ref{gaugespin2}), the gauge-invariant Lagrangian  $\mathcal{L}_{SZ}$ of Eq. (\ref{LSZ}) reduces to the Fierz-Pauli Lagrangian \cite{FierzP1939}, which was proposed to describe the massive spin-two particle 
\beq \label{LFP1}
{\cal L}_{FP} &=& \frac{1}{4} h_{\m \n}\,\p^2 \,h^{\m  \n} + \half (\p^\n h_{\m \n})^2
- \half \p^\n h_{\m\n} \p^\m h - \frac{1}{4} h \p^2 h - \frac{1}{4} \left(m^2 h^{\m\n} h_{\m\n}
- m^2 h^2 \right).
\eeq 
 For the massless case, $m=0$, the Fierz-Pauli Lagrangian is identified as the linearized Lagrangian of Einstein's gravity. 
Thus, the Fierz-Pauli Lagrangian corresponds to the gauge-invariant Lagrangian $\mathcal{L}_{FP}$ for a particular gauge. 
We may call this gauge condition, Eq. (\ref{FPgauge}), as the Fierz-Pauli gauge condition.

It may not be difficult to see that the Fierz-Pauli gauge condition Eq. (\ref{FPgauge}) fixes the gauge degrees of freedom
consistently. Consider the gauge orbits near the gauge fixing surface, $(h_{\m\n}, B_\m, \eta)$. Then, we may 
transform the fields such that they satisfy the Fierz-Pauli gauge condition: 
\begin{subequations}
\beq
0 &=& B_\m + \d B_\m = B_\m + \p_\m \e + m \e_\m, \\
0 &=& \eta + \frac{1}{2} h + \d \eta + \frac{1}{2}\d h =  \eta + \frac{1}{2} h - \frac{m}{4} (d-8) \e. 
\eeq 
\end{subequations}
These equations have a unique solution,
\begin{subequations}
\beq
\e &=& \frac{4}{m(d-8)}\left(\eta + \frac{1}{2} h \right), \\
\e_\m &=& - \frac{B_\m}{m} - \frac{4}{m^2 (d-8)} \left(\p_\m \eta + \half \p_\m h \right).
\eeq 
\end{subequations}
Therefore, the Fierz-Pauli gauge condition fixes the gauge degrees of freedom consistently and completely. 
 
\section{Canonical Quantization of the Fierz-Pauli Lagrangian}

Being a gauge fixed Lagrangian, the Fierz-Pauli Lagrangian is expected to possess second class constraints, similar to the Proca Lagrangian. 
Defining the canonical momenta for the Fierz-Pauli Lagrangian 
\beq
\Pi_{00} &=& \frac{\p {\cal L}_{FP}}{\p \dot h^{00}}, ~~ 
\Pi_{0i} = \frac{\p {\cal L}_{FP}}{\p \dot h^{0i}}, ~~ 
\Pi_{ij} = \frac{\p {\cal L}_{FP}}{\p \dot h^{ij}}, ~~
\eeq           
we find that
\begin{subequations}
\beq
\Pi_{00} &=&0,~~~\Pi_{0i}=\p^j h_{ij}-\p_i h^{jk}\eta_{jk}, \label{momenta1}\\
\Pi_{ij} &=& \half \dot{h}_{ij}-\half \eta_{ij}\dot{h}^{kl}\eta_{kl}.
\eeq
\end{subequations}
The first two equations of Eq. (\ref{momenta1}), which do not contain time derivatives of the canonical variables, are identified as the primary constraints:
\begin{subequations} 
\beq
\varphi_0 &=& \Pi_{00}  = 0, \\
\varphi_i &=& \Pi_{0i}-\p^j h_{ij}+\p_i h^{jk}\eta_{jk} . 
\eeq
\end{subequations}
The Hamiltonian corresponding to the Fierz-Pauli Lagrangian is found as
\begin{subequations}
\beq
H &=& \Pi_{0i}\dot{h}^{0i}+\Pi_{ij}\dot{h}^{ij}-\mathcal{L}_{FP} \nn\\
&=& (\Pi_{ij})^2-\frac{1}{d-2}(\Pi_{ij}\eta^{ij})^2+V, \\
V &=& -\half(\p_i h^{0j})^2+\frac{1}{4}(\p_k h^{ij})^2+\half (\p_i h^{0i})^2-\half(\p_i h^{ij})^2 \nn\\
&&-\half \p^i h^{00}\p^j h_{ij} +\half \p^j h_{ij}\p^i h^{kl}\eta_{kl}+\half \p_i h^{00}\p^i h^{jk}\eta_{jk}
+\frac{1}{4}(\p_i h^{jk}\eta_{jk})^2 \nn\\
&&+\frac{1}{4}m^2\left(-2(h^{0i})^2+(h^{ij})^2+2h^{00}h^{ij}\eta_{ij}-(h^{ij}\eta_{ij})^2\right).
\eeq 
\end{subequations}
Commutators of the primary constraints with the Hamiltonian generate the secondary constraints as
\begin{subequations}
\beq
\chi_0 &=& \left[ \varphi_0, H \right]_{PB} = -\half \p_i \p_j h^{ij}+\half \nabla^2h^{ij}\eta_{ij}-\half m^2 h^{ij}\eta_{ij}=0,  \\
\chi_i &=& \left[ \varphi_i, H \right]_{PB} =-\nabla^2 h^{0}{}_i+\p_i \p_j h^{0j}+m^2 h^{0}{}_i-2\p_j \Pi^j{}_i=0.
\eeq  
\end{subequations}

Using algebraic manipulations, we find that they are the second class constraints for $m \not=0$:
\begin{subequations}
\beq
\left[\varphi_0(\bolx), \varphi_i(\bolx^\prime) \right]_{PB} &=& 0, ~~~\left[\varphi_0(\bolx), \chi_0(\bolx^\prime) \right]_{PB} =0,  \\
\left[\varphi_0(\bolx), \chi_i(\bolx^\prime) \right]_{PB} &=& 0, ~~~\left[\varphi_i(\bolx), \varphi_j(\bolx^\prime) \right]_{PB} =0,  \\
\left[\varphi_i(\bolx), \chi_0(\bolx^\prime) \right]_{PB} &=& 0, ~~~\left[\varphi_i(\bolx), \chi_j(\bolx^\prime) \right]_{PB} =-m^2 \d_{ij} 
\d(\bolx -\bolx^\prime),  \\
\left[\chi_i(\bolx), \chi_j(\bolx^\prime) \right]_{PB} &=&0, ~~~\left[\chi_0(\bolx), \chi_i(\bolx^\prime) \right]_{PB} = -m^2 \p_i \d(\bolx -\bolx^\prime).
\eeq 
\end{subequations}
Therefore, we should employ the Dirac brackets to perform the canonical quantization of the Fierz-Pauli model, which is 
expected to be very complicated. Moreover, as we observed in the previous section on the discussion of the Proca model, 
the Dirac brackets may not be well defined in the massless limit. This leads us to conclude that the Fierz-Pauli Lagrangian may not 
be suitable for studying the small-mass limit of the massive spin-two particle. We should therefore look for an alternative Lagrangian 
that is well defined in the small-mass limit yet gauge equivalent to the Fierz-Pauli Lagrangian if we are interested in 
studying the spin-two particle with a small mass. 

Although it is cumbersome to carry out canonical quantizations of the Fierz-Pauli Lagrangian, it is not difficult to
find the propagator of the massive spin-two particle described by the Fierz-Pauli Lagrangian: 
In momentum space, we may rewrite the Fierz-Pauli Lagrangian of Eq. (\ref{LFP1}) as 
\begin{subequations}
\beq
{\cal L}_{FP} &=& \frac{1}{2} h^{\m\n} {\cal O}^{FP}_{\m\n,\a\b} h^{\a\b}, \\
{\cal O}^{FP}_{\m\n,\a\b} &=& -\frac{(p^2+m^2)}{4}( \eta_{\m \a}\eta_{\n \b}+ \eta_{\n \a}\eta_{\m \b}-2\eta_{\m\n}\eta_{\a\b})
-\half (\eta_{\m\n}p_\a p_\b+ \eta_{\a\b}p_\m p_\n) \nn\\
&&+\frac{1}{4}(\eta_{\m\a}p_\n p_\b+\eta_{\m\b}p_\n p_\a+\eta_{\n\a}p_\m p_\b+\eta_{\n\b}p_\m p_\a).
\eeq 
\end{subequations}
The propagator of the spin-two field is defined by 
\beq
\mathcal{O}^{FP}_{\m \n, \a \b}G^{\a \b, \s \l}_{FP}=\frac{i}{2}\left(\d^\s_\m \d^\l_\n+\d^\s_\n \d^\l_\m\right).
\eeq
Using algebraic manipulations, we obtain 
\beq \label{propagatorfp}
G^{\a \b, \s \l}_{FP} &=& \frac{i}{p^2+m^2} \biggl\{\frac{2}{d-1}\left(\eta_{\a\b}+\frac{p_\a p_\b}{m^2}\right)\left(\eta_{\s\l}+\frac{p_\s p_\l}{m^2}\right) \nn\\
&&-\left(\eta_{\a\s}+\frac{p_\a p_\s}{m^2}\right)\left(\eta_{\b\l}+\frac{p_\b p_\l}{m^2}\right)
-\left(\eta_{\a\l}+\frac{p_\a p_\l}{m^2}\right)\left(\eta_{\b\s}+\frac{p_\b p_\s}{m^2}\right)
\biggr\}.
\eeq 
This propagator of the spin-two field for the Fierz-Pauli Lagrangian and the propagator of the vector field Eq. (\ref{procagauge})
for the Proca model are similar in that both diverge in the massless limit.  Therefore, the Fierz-Pauli Lagrangian may not be useful for studying the spin-two particle with small mass. 


\section{Canonical Quantization of Massive Symmetric Rank-Two Tensor} 

Returning to the gauge-invariant Lagrangian of Siegel and Zwiebach, we begin the canonical quantization of the massive symmetric 
rank-two tensor by defining the canonical momenta: We may rewrite ${\cal L}_{SZ}$ of Eq. (\ref{LSZ}) as follows
\beq \label{LSZt}
\mathcal{L}_{SZ} &=& -\frac{1}{4}(\dot{h}^{00})^2+\dot{h}^{00}(-\p_i h^{0i}+\dot{\eta}+mB^0)+\dot{h}^{0i}(\p^j h_{ij}+\p_i \eta-mB_i) \nn\\
&&+\frac{1}{4}(\dot{h}^{ij})^2+\dot{B}^0(\p_i B^i+\frac{1}{4}m(-h^{00}+h^{ij}\eta_{ij})+\frac{5}{2}m\eta)+\half (\dot{B}^i)^2 \nn\\
&&- \dot{\eta}^2+\p_i h^{0i}\dot{\eta}-V.
\eeq
Here,  $V$ denotes a potential term that does not contain time derivatives of the fields, 
\beq \label{LSZV}
V &=& \frac{1}{4}(\p_i h^{00})^2+\frac{7}{32}m^2(h^{00})^2-\half(\p_i h^{0j})^2+\half (\p_i h^{0i})^2-\half m^2(h^{0i})^2 \nn\\
&& +\frac{1}{4}(\p_k h^{ij})^2-\half(\p_i h^{ij})^2+\frac{1}{4}m^2(h^{ij})^2-\frac{1}{32}m^2(h^{ij}\eta_{ij})^2
-\half(\p_i B^0)^2 \nn\\
&& +\half(\p_i B^j)^2-\half(\p_i B^i)^2-(\p_i \eta)^2-\frac{13}{8}m^2\eta^2+\frac{1}{4}mh^{00}\p_i B^i
+\frac{1}{16}m^2h^{00}h^{ij}\eta_{ij} \nn\\
&&+\frac{3}{8}m^2\eta h^{00}-m\p_i h^{0i}B^0-\p_i h^{ij}\p_j \eta+m\p_i h^{ij}B_j-\frac{1}{4}mh^{ij}\eta_{ij}\p_k B^k \nn\\
&&-\frac{3}{8}m^2\eta h^{ij}\eta_{ij}-\frac{5}{2}m\eta \p_i B^i .
\eeq 
The canonical momenta that are conjugate to the fields are defined as 
\begin{subequations} 
\beq
\Pi_{00} &=& \frac{\p \mathcal{L}_{SZ}}{\p \dot{h}^{00}}, ~~ \Pi_{0i} = \frac{\p \mathcal{L}_{SZ}}{\p \dot{h}^{0i}},~~
\Pi_{ij}=\frac{\p \mathcal{L}_{SZ}}{\p \dot{h}^{ij}},  \label{LSZm1}\\
\Pi^B_{0} &=& \frac{\p \mathcal{L}_{SZ}}{\p \dot{B}^{0}},~~\Pi^B_{i}=\frac{\p \mathcal{L}_{SZ}}{\p \dot{B}^{i}},~~
\pi_{\eta}=\frac{\p \mathcal{L}_{SZ}}{\p \dot{\eta}}. \label{LSZm2}
\eeq
\end{subequations}
From Eqs. (\ref{LSZt}, \ref{LSZV}, \ref{LSZm1}, \ref{LSZm2}) we obtain 
\begin{subequations}
\beq
\Pi_{00} &=&-\half \dot{h}^{00}-\p_i h^{0i}+\dot{\eta}+mB^0,~~~
\Pi_{0i}=\p^j h_{ij}+\p_i \eta-mB_i,   \label{mom1}\\
\Pi_{ij} &=& \half \dot{h}_{ij},~~~~ \Pi^B_{0}=\p_i B^i+\frac{1}{4}m(-h^{00}+h^{ij}\eta_{ij})+\frac{5}{2}m\eta, \label{mom2}\\
\Pi^B_{i} &=& \dot{B}_i, ~~~~~~~ \pi_{\eta} = \dot{h}^{00}-2\dot{\eta}+\p_i h^{0i} \label{mom3}
\eeq
\end{subequations}
and the Hamiltonian density
\beq
{\cal H}_{SZ} 
&=& -\Pi_{00}^2+\Pi_{ij}^2+\half \left(\Pi^B_{i}\right)^2-\pi_{\eta}(mB^0-\p_i h^{0i}) \nn\\
&& +mB^0(mB^0-\p_i h^{0i})+V.
\eeq

We may manufacture the primary constraints by taking the linear combination of equations Eqs.(\ref{mom1}, \ref{mom2}, \ref{mom3}):
\begin{subequations}
\beq
\varphi_0 &=& 2\Pi_{00}+\pi_{\eta}+\p_i h^{0i}-2mB^0=0, \label{primea}\\
\varphi_i &=&\Pi_{0i}-\p^j h_{ij}-\p_i \eta+mB_i=0, \label{primeb}\\
\varphi^B &=& \Pi^B_{0}-\p_i B^i+\frac{1}{4}m(h^{00}-h^{ij}\eta_{ij})-\frac{5}{2}m\eta=0. \label{primec}
\eeq
\end{subequations}
Commutators of these primary constraints with the Hamiltonian yield the secondary constraints as follows: 
\begin{subequations}
\beq
\chi_0 &=& [\varphi_0, H_{SZ}]_{PB}=\nabla^2 h^{00}-m^2h^{00}+m\pi_{B0}-\p^i \pi_{0i}-\nabla^2 \eta, \label{seconda}\\
\chi_i &=& [\varphi_i, H_{SZ}]_{PB}=\p_i \pi_{\eta}-m\p_i B^0-\nabla^2 h^{0i}+m^2 h^{0i}+m\pi_{Bi}-2\p^j \pi_{ij}, 
\label{secondb}\\
\chi^B &=& [\varphi^B ,H_{SZ}]_{PB}=\half m\pi_{00}+\frac{3}{2}m\pi_{\eta}-\p^i \pi_{Bi}-\half m \eta_{ij}\pi^{ij}-\half m^2B^0-\nabla^2 B^0, \label{secondc}
\eeq
\end{subequations}
where $H_{SZ} = \int d^{d-1} \, \bolx {\cal H}_{SZ}(\bolx)$. The primary constraints, Eqs. (\ref{primea}, \ref{primeb}, \ref{primec}), and the secondary constraints, Eqs. (\ref{seconda}, \ref{secondb}, \ref{secondc}), form a set of firs class constraints, which generate the local gauge transformation in Eq.(\ref{gaugespin2}) 
\beq
\d h_{\m\n} &=& \left[h_{\m\n}, \Omega{(\e^0, \e^i, \e)}\right]_{PB}, ~~ \d B_\m = \left[B_\m, \Omega{(\e^0, \e^i, \e)}\right]_{PB}, ~~
\d \eta = \left[\eta, \Omega{(\e^0, \e^i, \e)}\right]_{PB}
\eeq 
with a gauge generator
\beq
\Omega{(\e^0, \e^i, \e)} &=& -\dot{\e}^0\varphi_0-\dot{\e}\varphi^B-\dot{\e}^i\varphi_i+\e^0\chi_0+\e \chi^B+\e^i \chi_i.
\eeq 

\section{Massive Symmetric Rank-Two Tensor in the Transverse-Traceless Gauge} 

We have shown that the gauge degrees of freedom of the Siegel-Zwiebach Lagrangian can be fixed by the Fierz-Pauli gauge condition, and the 
Lagrangian reduces to the well-known Fierz-Pauli Lagrangian, which describes a massive spin-two particle. However, the Lagrangian in the Fierz-Pauli gauge may not be useful for studying a spin-two particle with small mass because the propagator for the spin-two field may be singular in the massless limit. From the study of the Proca model, we deduced that an appropriate gauge condition may be the transverse-traceless (TT) gauge \cite{Gravitation1973,Brandt2007,Gribosky1989,Rebhan1991,Blasi2017} 
\beq
\p_\m h^\m{}_\n =0, ~~~ h = 0 . 
\eeq 
In the TT gauge, ${\cal L}_{SZ}$ reduces to ${\cal L}_{\rm TT}$:
\beq\label{LagTT}
\mathcal{L}_{\rm TT} &=& \frac{1}{4}h_{\m \n}(\p^2-m^2)h^{\m  \n}+\half B_\m (\p^2 -m^2)B^\m -\half \eta (\p^2-m^2)\eta \nn\\
&& +\half (\p_\m \eta -mB_\m)^2+\half(\frac{3}{2}m\eta+\p_\m B^\m)^2 -  \frac{\l}{2} \left(\p_\m h^\m{}_\n \right)^2 -
\frac{\s}{2} h^2 
\eeq 
where $\l$ and $\s$ are Lagrangian multipliers. We note that the symmetric rank-two tensor field $h_{\m\n}$ decouples 
from the Stueckelberg fields $B_\m$ and $\eta$. Thus, we can separate the propagator for $h_{\m\n}$ from those of $B_\m$ and $\eta$. 

It takes some algebraic manipulation to obtain the propagator for $h_{\m\n}$. In momentum space, the Lagrangian for $h_{\m\n}$ in the TT gauge may be written as 
\begin{subequations}
\beq
{\cal L}_{TT}^h &=& \half h^{\m \n}\mathcal{O}_{\m \n,\a \b}h^{\a \b}, \\
\mathcal{O}_{\m \n,\a \b}&=&-\frac{(p^2+m^2)}{4}(\eta_{\m \a}\eta_{\n \b}+\eta_{\n \a}\eta_{\m \b})
- \s \, \eta_{\m \n}\eta_{\a \b} \nn\\
&& -\frac{\l}{4}(\eta_{\m \a}p_\n p_\b+\eta_{\m \b}p_\n p_\a+\eta_{\n \a}p_\m p_\b+\eta_{\n \b}p_\m p_\a) ,
\eeq 
\end{subequations}
and the propagator for $h_{\m\n}$, $G^{\a \b, \s \l}$, which satisfies 
\beq
\mathcal{O}_{\m \n, \a \b}G^{\a \b, \s \l}=\frac{i}{2}\left(\d^\s_\m \d^\l_\n+\d^\s_\n \d^\l_\m\right).
\eeq
Some straightforward but tedious algebra yields an explicit form of the propagator for $h_{\m\n}$ in the TT gauge: 
\begin{subequations}
\beq
G^{\a \b, \s \l}&=& \frac{i}{p^2+m^2}\biggl\{A\eta^{\a \b}\eta^{\s \l}-(\eta^{\a \s}\eta^{\b \l}+\eta^{\a \l}\eta^{\b \s})
+\frac{C}{p^2}(\eta^{\a \b}p^\s p^\l+\eta^{\s \l}p^\a p^\b) \nn\\
&&+\frac{D}{p^2}(\eta^{\a \s}p^\b p^\l+\eta^{\a \l}p^\b p^\s+\eta^{\b \s}p^\a p^\l+\eta^{\b \l}p^\a p^\s)+\frac{E}{p^4}p^\a p^\b p^\s p^\l \biggr\}, \\
A &=& \frac{4 p^2 \left(p^2 + m^2 \right) \l + 8 p^2 \l \s}{\left(p^2 +m^2\right)^2 + 2\left(p^2+m^2\right) \left(p^2 \l + d\s \right)
 + 4(d-1) p^2 \l \s} ,\\
C &=& - \frac{8 p^2 \l \s}{\left(p^2 +m^2\right)^2 + 2\left(p^2+m^2\right) \left(p^2 \l + d\s \right)
 + 4(d-1) p^2 \l \s},\\
D &=& \frac{p^2 \l}{\left(p^2 + m^2 \right) + p^2 \l }, \\
E &=& - \frac{4 p^2 \l}{\left(p^2 + m^2 \right) + p^2 \l } \frac{\left(p^2+m^2\right)p^2 \l + 2(d-2) p^2 \l \s}{\left(p^2 +m^2\right)^2 + 2\left(p^2+m^2\right) \left(p^2 \l + d\s \right) + 4(d-1) p^2 \l \s} .
\eeq
\end{subequations}
In the limit where $\l, \s \rightarrow \infty$, we find 
\beq
A \rightarrow \frac{2}{d-1}, ~~~ C \rightarrow - \frac{2}{d-1}, ~~~ D \rightarrow 1, ~~~
E \rightarrow - \frac{2(d-2)}{d-1}. 
\eeq 
Hence, we obtain the propagator, $G_{TT}^{\a \b, \s \l}(m)$ in the TT gauge, which reduces to
\beq \label{propaTT}
\lim_{\l, \s \rightarrow \infty} G_{TT}^{\a \b, \s \l} (m)
&=& \frac{i}{p^2+m^2}\Biggl\{ \frac{2}{d-1} \left(\eta^{\a\b} - \frac{p^\a p^\b}{p^2} \right)
\left(\eta^{\s\l} - \frac{p^\s p^\l}{p^2} \right) \nn\\
&& - \left(\eta^{\a\s} - \frac{p^\a p^\s}{p^2} \right) \left(\eta^{\b\l} - \frac{p^\b p^\l}{p^2} \right)
- \left(\eta^{\a\l} - \frac{p^\a p^\l}{p^2} \right) \left(\eta^{\b\s} - \frac{p^\b p^\s}{p^2} \right)
\Biggr\}
\eeq
in the limit $\l, \s \rightarrow \infty$. It is easy to check that $G_{TT}^{\a \b, \s \l}(m)$ satisfies the transverse-traceless 
gauge condition:
\begin{subequations}
\beq
G_{TT}^{\a \b, \s \l} (m) p_\a &=& G_{TT}^{\a \b, \s \l}(m)p_\b = G_{TT}^{\a \b, \s \l} (m)p_\s = G_{TT}^{\a \b, \s \l} (m)p_\l = 0, \\
\eta_{\a\b} G_{TT}^{\a \b, \s \l}(m) &=& \eta_{\s\l} G_{TT}^{\a \b, \s \l} (m)=0. 
\eeq
\end{subequations}
Apparently, the propagator for the massive spin-two field in the TT gauge has a well-defined massless limit: 
\beq
\lim_{m \rightarrow 0} G_{TT}^{\a \b, \s \l} (m)  &=&  G_{TT}^{\a \b, \s \l} (0)  \nn\\
&=& \frac{i}{p^2}\Biggl\{ \frac{2}{d-1} \left(\eta^{\a\b} - \frac{p^\a p^\b}{p^2} \right)
\left(\eta^{\s\l} - \frac{p^\s p^\l}{p^2} \right) \nn\\
&& - \left(\eta^{\a\s} - \frac{p^\a p^\s}{p^2} \right) \left(\eta^{\b\l} - \frac{p^\b p^\l}{p^2} \right)
- \left(\eta^{\a\l} - \frac{p^\a p^\l}{p^2} \right) \left(\eta^{\b\s} - \frac{p^\b p^\s}{p^2} \right)
\Biggr\}.
\eeq 
Here, $G_{TT}^{\a \b, \s \l} (0)$ is identified as the propagator for the massless graviton in the TT gauge \cite{Gravitation1973}.

\begin{figure}[htbp]
\begin {center}
\epsfxsize=0.8\hsize

\epsfbox{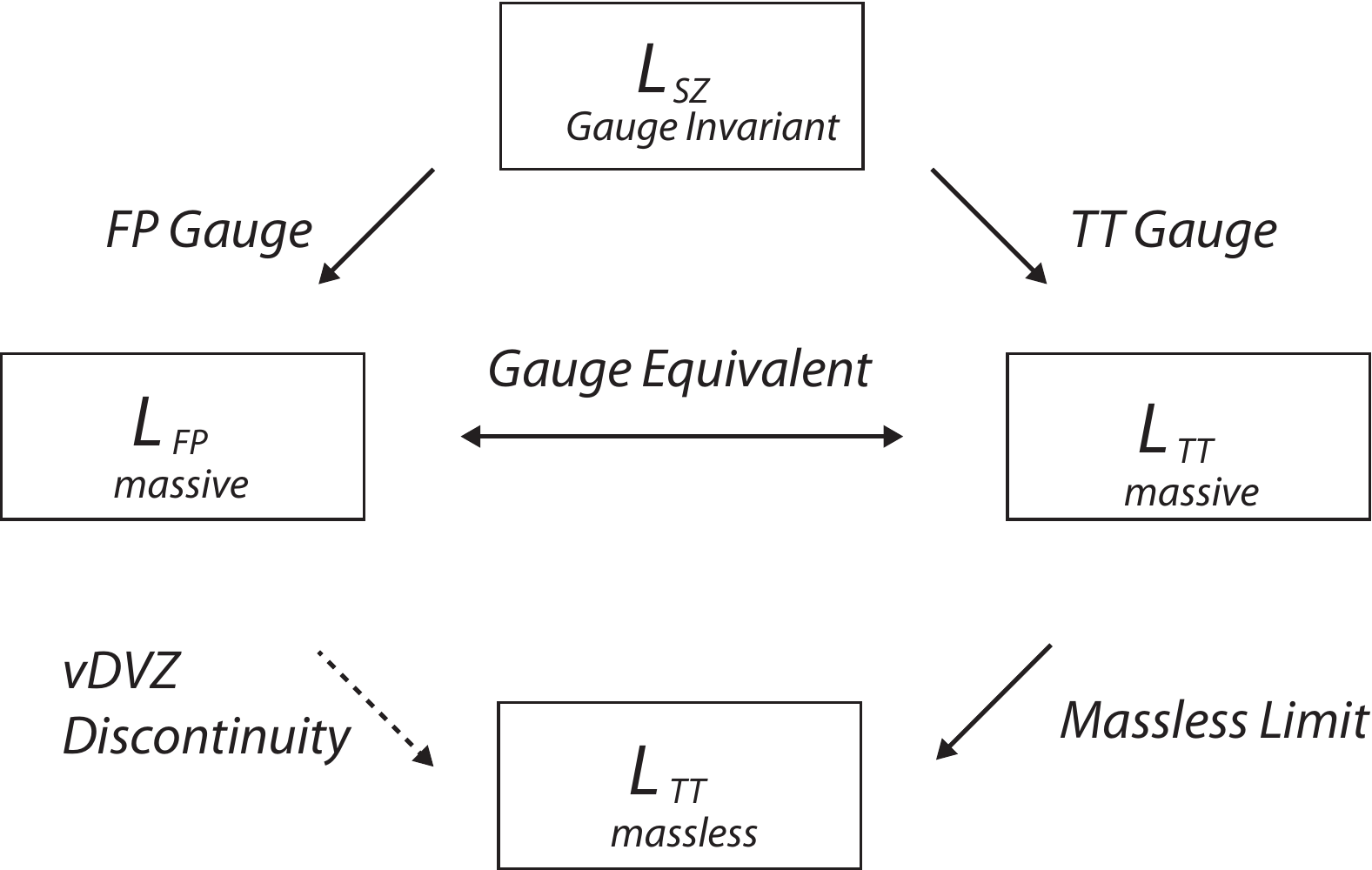}
\end {center}
\caption {\label{spintwo} Relations between theories of the spin-two field.}
\end{figure}

Returning to the Lagrangian in the TT gauge of Eq. (\ref{LagTT}), we may write the part of the Lagrangian containing the Stueckelberg fields $B_\m$ and $\eta$ as 
\beq\label{LagB}
{\cal L}^{B,\eta}_{TT} &=& \frac{1}{4} \left(\p_\m B_\n -\p_\n B_\m \right)^2 + \frac{5}{2} m \eta \p_\m B^\m  - \frac{1}{2} \eta\left(\p^2 -m^2 \right) \eta .  
\eeq 
We note that the kinetic term for $B_\m$ is invariant under $U(1)$ gauge transformation
\beq
B_\m \rightarrow B_\m + \p_\m \Lambda. 
\eeq 
Decomposing the vector field $B_\n$ into a transvers part $B^T_\n$ and a longitudinal part $B^L_\n$
\beq
B_\n = B^T_\n + B^L_\n = B^\m \left(\eta_{\m\n} - \frac{p_\m p_\n}{p^2} \right) + B^\m \left(\frac{p_\m p_\n}{p^2} \right),
\eeq 
we find that $\eta$ couples to the longitudinal part of the $B_\m$ field, which may be written with a scalar field $\phi$ 
as $B^L_\m = \p_\m \phi$. 
As $B^L$, equivalently $\phi$, appears only in the Lagrangian Eq. (\ref{LagB}) through the linear coupling to $\eta$, $\frac{5}{2} m \p^2 \eta \, \phi$ , $\phi$ plays the role of a Lagrangian multiplier. Integrating $\phi$ in the path integral imposes a condition
$\p^2 \eta =0$: therefore, $\eta$ is an auxiliary field, which can be trivially integrated. Hence, integrating $B^L_\m$ and $\eta$, we get ${\cal L}^{B,\eta}_{TT} \rightarrow {\cal L}^{B}_{TT}$, which may be written by
\beq
{\cal L}^B_{TT} = \frac{1}{4} \left(\p_\m B_\n -\p_\n B_\m \right)^2, ~~~ \p_\m B^\m = 0. 
\eeq 
The propagator for $B^T_\m$ is the bona-fide propagator for a massless $U(1)$ gauge field in the Lorenz gauge
\beq
G^B_{\m\n} = \frac{i}{p^2}\left(\eta_{\m \n} - \frac{p_\m p_\n}{p^2}\right), ~~~ p^\m G^B_{\m\n} = p^\n G^B_{\m\n} = 0. 
\eeq 
Collecting ${\cal L}^h_{TT}$ and ${\cal L}^B_{TT}$, we find that the Siegel-Zwiebach model for a massive 
symmetric rank-two tensor is described in the TT gauge by a massive spin-two field, satisfying the TT gauge condition and 
the Stueckelberg vector field $B_\m$, which becomes a massless $U(1)$ gauge field, satisfying the Lorenz gauge condition
\beq \label{LTT2}
{\cal L}_{TT} &=& \frac{1}{4}h_{\m \n}(\p^2-m^2)h^{\m  \n} + \frac{1}{4} \left(\p_\m B_\n -\p_\n B_\m \right)^2
-  \frac{\l}{2} \left(\p_\m h^\m{}_\n \right)^2 -\frac{\s}{2} h^2 - \frac{\a}{2} \left(\p_\m B^\m \right)^2. 
\eeq  
It may be reasonable to integrate the Stueckelberg vector field $B_\m$ further because it does not couple
to physical degrees of freedom: in conclusion, we may describe the massive spin-two particle by ${\cal L}^h_{TT}$ 
\beq \label{LhTT2}
{\cal L}^h_{TT} &=& \frac{1}{4}h_{\m \n}(\p^2-m^2)h^{\m  \n} -  \frac{\l}{2} \left(\p_\m h^\m{}_\n \right)^2 -\frac{\s}{2} h^2 
\eeq 
and the propagator $G_{TT}^{\a \b, \s \l}(m)$ in the TT gauge given by Eq. (\ref{propaTT}). Fig. \ref{spintwo} summarizes the relations between the theories of the spin-two field. 

\section{Discussions and Conclusions}

We have studied the massive symmetric rank-two tensor in open string theory in the framework of canonical quantization
that contains two Stueckelberg fields. Performing canonical analysis, we find that the massive symmetric rank-two tensor 
theory of Siegel and Zwiebach contain only first class constraints. By explicit calculation, we have shown that
the rank-two tensor theory is invariant under the local gauge transformation generated by the first class constraints in the
critical dimension of bosonic string theory. In a particular gauge, the Stueckelberg fields the rank-two 
tensor theory reduces to the massive spin-two theory of Fierz and Pauli. Then, we pointed out that the massive spin-two theory
of Fierz and Pauli possesses second class constraints, and the Dirac brackets are not well defined in the massless limit.
Accordingly, the propagator for the massive spin-two field of Fierz-Pauli theory also diverges and does not reduce to 
that of the massless graviton of Einstein's gravity in the massless limit: this gives rise to the vDVZ discontinuity 
\cite{vanDam1970,Zakharov1970}, which has been studied in detail in Refs. \cite{Vainshtein1972,Porrati2002,Faria2017,Myung2017}. 

Because the massive symmetric rank-two tensor theory is gauge invariant, we may choose a gauge condition alternative to the 
Fierz-Pauli gauge condition. We may impose the gauge fixing condition on the spin-two field $h_{\m\n}$ instead of the 
Stueckelberg fields: the most plausible choice may be the transverse-traceless (TT) gauge condition by which the spin-two 
field completely decouples from the two Stueckelberg fields. Hence, in the TT gauge, 
we may integrate the Stueckelberg fields, which may not couple to physical sources, to obtain a simple 
Lagrangian ${\cal L}^h_{TT}$, Eq. (\ref{LhTT2}) describing the massive spin-two field satisfying the TT gauge condition. 
By some algebra, we constructed a propagator for massive spin-two field, $G_{TT}^{\a \b, \s \l}(m)$ in the TT gauge
In contrast to the propagator in the massive spin-two theory of Fierz and Pauli, $G^{\a \b, \s \l}_{FP}(m)$, Eq.(\ref{propagatorfp}), 
the propagator in the TT gauge, $G_{TT}^{\a \b, \s \l}(m)$ smoothly reduces in the zero-mass limit to the graviton propagator of Einstein's gravity in the TT gauge without encountering any discontinuity. 
It may be interesting to study this massive rank-two
tensor theory further to understand how the Vainshtein mechanism \cite{Vainshtein1972} is realized in the TT gauge. 
It may also be worthwhile to investigate the propagation of the gravitational wave using the propagator in the TT gauge in details
to  explore the possibility of finite mass for graviton. It is apparent that the massive rank-two tensor theory of Siegel and Zwiebach 
is free of Boulware-Deser ghost \cite{BoulwareD1972,BoulwareD1975PL,BoulwareD1975NP}, because this gauge-invariant theory reduces to the Fierz-Pauli theory in a particular gauge, where the Boulware-Deser ghost is absent. 

As we have shown by explicit calculation, the symmetric rank-two tensor theory of Siegel and Zwiebach possesses local gauge
invariance only at the critical dimension. In order to apply this symmetric rank-two tensor theory on four-dimensional space-time,
we need to define the open string theory on a $D3$-brane. In this case we expect to have  superfluous vector and scalar fields. 
The present work may be extended further to symmetric rank-two tensor theories on various dimensions by defining open string 
theories on general $Dp$-branes. The symmetric rank-two tensor theory based on string theory is also useful to 
construct a consistent interacting spin-two field theory \cite{VeloZa1969,VeloZb1969,Velo1972,Argyres1989,Porratia2011}, which is one of most challenging problems in theoretical physics. We can obtain cubic coupling terms directly from three open string 
scattering amplitudes, evaluating Polyakov string path integrals \cite{Polyakov1981} in the proper-time gauge 
\cite{TLeeann1988,TLeePLB2017,TLeeJKPS2017,Lee2017d,TLee2017cov,TLee2019PL}. 
Choosing external string states as a composition of vector gauge field and spin-two field, we may fix the cubic interaction
terms between vector gauge field and spin-two field. In principle, string scattering amplitudes will guide us to construct a consistent interacting spin-two field theory and we may resolve the Velo-Zwanziger problem \cite{VeloZa1969,VeloZb1969,Velo1972} in the framework of string theory on $Dp$-branes. The present work may be extended to describe interaction between massive 
spin-two field with massless graviton, introducing splitting and joining interactions between open and closed strings.
Evaluating string scattering amplitudes of open and closed strings, we may obtain interaction terms between the massive 
spin-two field of open string and massless graviton of closed string. The resultant may provide an alternative approach to
the problem of consistent equations of motion which has been studied in Refs. \cite{Buchbinder1999NP,Buchbinder1999PLB}.
Extensions of the present work along these directions will be presented elsewhere. 

It is interesting to note that the massive symmetric rank-two tensor appears also in the spectrum of closed string if we choose a left-right asymmetric 
vacuum \cite{KLee2017JHEP}. It has been shown that with asymmetric vacuum and an appropriately chosen time ordering, the spectrum of closed string contains only a finite number of field theoretic degrees of freedom: string gravity fields, which consists of massless spin-two graviton, Kalb-Ramond and dilaton fields, and two massive symmetric rank-two tensor fields. We expect that the BRST invariant formulation of this closed string theory may introduce additional Stueckelberg fields as in the case of open string theory and the present work may be useful to study the canonical structure of this theory.


\vskip 1cm

\begin{acknowledgments}
This work was supported by the Basic Science Research Program through the National Research Foundation of Korea (NRF) funded by the Ministry of Education (No. 2017R1D1A1A02017805). 
We thank Soo-Jong Rey and Kanghoon Lee for informing us of their recent work and comments. 
TL acknowledges the hospitality at APCTP, where a part of this work was done.  
\end{acknowledgments}


\appendix

\section{Gauge Symmetry of Massive Symmetric Rank-Two Tensor Theory}

It may be instructive to check the local gauge invariance of the massive symmetric rank-two tensor theory $\mathcal{L}_{SZ}$
of Siegel and Zwiebach by explicit calculations. The massive symmetric rank-two tensor theory of Siegel and Zwiebach is
described by the following Lagrangian:
\beq 
\mathcal{L}_{SZ} &=& \frac{1}{4}h_{\m \n}(\p^2-m^2)h^{\m  \n}+\half B_\m (\p^2 -m^2)B^\m -\half \eta (\p^2-m^2)\eta \nn\\
&& +\half (\p^\n h_{\m \n}+\p_\m \eta -mB_\m)^2+\half(\frac{m}{4}h+\frac{3}{2}m\eta+\p_\m B^\m)^2
\eeq
where $h = h^\s{}_\s$. From the BRST invariance \cite{SiegelZ1986}, we deduce the gauge transformation as
\beq 
\d h_{\m \n} = \p_\m \e_\n + \p_\n \e_\m -\half m \eta_{\m \n}\e,~~~
\d B_\m = \p_\m \e + m\e_\m,~~~\d \eta = -\p_\m \e^\m+\frac{3}{2}m\e . 
\eeq
Under gauge transformation, each term in $\mathcal{L}_{SZ}$ is transformed as follows: 
\begin{subequations}
\beq
\frac{1}{4}h_{\m \n}\p^2h^{\m \n} &\to&
\half \p_\m \e_\n \p^2 \p^\m \e^\n+\half \p_\m \e_\n \p^2 \p^\n \e^\m+\frac{1}{16}m^2d\e \p^2 \e \nn\\
&&+ h_{\m \n}\p^2 \p^\m \e^\n-\frac{1}{4}m h \p^2 \e -\half m \p_\m \e^\m \p^2 \e, \\
-\frac{m^2}{4}h_{\m \n}h^{\m \n} &\to&
-\half m^2 \p_\m \e_\n \p^\m \e^\n-\half m^2\p_\m \e_\n \p^\n \e^\m-\frac{1}{16}m^4 d\e^2 \nn\\
&&-m^2h_{\m \n}\p^\m \e^\n+\frac{1}{4}m^3h\e+\half m^3 \p_\m \e^\m \e, \\
\half B_\m \p^2 B^\m &\to&
\half \p_\m \e \p^2 \p^\m \e+\half m^2 \e_\m \p^2 \e^\m+B_\m \p^2 \p^\m \e \nn\\
&&+mB_\m \p^2 \e^\m+m\p_\m \e \p^2 \e^\m, \\
-\eta \p^2 \eta &\to&
-\p_\m \e^\m \p^2 \p_\n \e^\n-\frac{9}{4}m^2\e \p^2 \e+2\eta \p^2 \p_\m \e^\m \nn\\
&&-3m\eta \p^2 \e+3m\p_\m \e^\m \p^2 \e, 
\eeq
\beq
\frac{13}{8}m^2 \eta^2 &\to&
\frac{13}{8}m^2\p_\m \e^\m \p_\n \e^\n+\frac{117}{32}m^4\e^2-\frac{13}{4}m^2\p_\m \e^\m \eta \nn\\
&&+\frac{39}{8}m^3\eta \e-\frac{39}{8}m^3\p_\m \e^\m \e, \\
\half \p^\n h_{\m \n}\p_\l h^{\m \l} 
&\to& 
-\half \p_\m \e_\n \p^\m \p^\n \p_\l \e^\l-\half \p_\n \e_\m \p^\n \p^2 \e^\m-\frac{1}{8}m^2\e \p^2 \e \nn\\
&&-h_{\m \n}\p^\m \p^\n \p_\l \e^\l-h_{\m \n}\p^\n \p^2 \e^\m+\half m h_{\m \n}\p^\m \p^\n \e \nn\\
&&-\p_\m \e_\n \p^\n \p^2 \e^\m+m\e \p^2 \p_\m \e^\m, \\
\p^\n h_{\m \n}\p^\m \eta 
&\to& h_{\m \n}\p^\m \p^\n \p_\l \e^\l
-\frac{3}{2}mh_{\m \n}\p^\m \p^\n \e-\p_\m \e_\n \p^\m \p^\n \eta \nn\\
&&+\p_\m \e_\n \p^\m \p^\n \p_\l \e^\l-\frac{3}{2}m\p_\m \e_\n \p^\m \p^\n \e-\p_\n \e_\m \p^\m \p^\n \eta \nn\\
&&+\p_\n \e_\m \p^\m \p^\n \p_\l \e^\l-\frac{3}{2}m\p_\n \e_\m \p^\m \p^\n \e+\half m\e \p^2 \eta \nn\\
&&-\half m \e \p^2 \p_\m \e^\m+\frac{3}{4}m^2\e \p^2 \e ,
\eeq
\beq
-m\p^\n h_{\m \n}B^\m 
&\to& mh_{\m \n}\p^\m \p^\n \e+m^2h_{\m \n}\p^\n \e^\m
+m\p_\m \e_\n \p^\n B^\m \nn\\
&&+m\p_\m \e_\n \p^\m \p^\n \e+m^2\p_\m \e_\n \p^\n \e^\m+m\p_\n \e_\m \p^\n B^\m \nn\\
&&+m\p_\m \e_\n \p^\m \p^\n \e+m^2\p_\n \e_\m \p^\n \e^\m-\half m^2\e \p_\m B^\m \nn\\
&&-\half m^2\e \p^2 \e-\half m^3 \e \p_\m \e^\m, \\
m\eta \p_\m B^\m &\to& m\eta \p^2 \e+m^2\eta \p_\m \e^\m-m\p_\n \e^\n \p_\m B^\m \nn\\
&&-m\p_\n \e^\n \p^2 \e-m^2\p_\n \e^\n \p_\m \e^\m+\frac{3}{2}m^2\e \p_\m B^\m \nn\\
&&+\frac{3}{2}m^2\e \p^2 \e+\frac{3}{2}m^3\e \p_\m \e^\m, \\
\frac{m^2}{32}h^\m_\n h^\n_\n &\to& \frac{1}{8}m^2\p_\m \e^\m \p_\n \e^\n+\frac{1}{128}m^4 d^2\e^2
+\frac{1}{8}m^2h\p_\n \e^\n \nn\\
&&-\frac{d}{32}m^3 h\e-\frac{1}{16}m^3d \p_\m \e^\m \e, 
\eeq
\beq
\half \p_\m B^\m \p_\n B^\n 
&\to& 
-\half \p^\m \e \p_\m \p^2 \e-\half m^2\e^\m \p_\m \p_\n \e^\n-B^\m \p_\m \p^2 \e \nn\\
&&-mB^\m \p_\m \p_\n \e^\n-m\e^\m \p_\m \p^2 \e, \\
\frac{3}{8}m^2 h^\m_\m \eta &\to& -\frac{3}{8}m^2h\p_\n \e^\n+\frac{9}{16}m^3h\e+\frac{3}{4}m^2\p_\m \e^\m \eta \nn\\
&&-\frac{3}{4}m^2\p_\m \e^\m \p_\n \e^\n+\frac{9}{8}m^3\p_\m \e^\m \e-\frac{3d}{16} m^3\e \eta \nn\\
&&+\frac{3d}{16} m^3\e \p_\m \e^\m-\frac{9}{32}m^4d\e^2 \\
\frac{1}{4}mh^\m_\m \p_\n B^\n &\to& \frac{1}{4}mh\p^2 \e+\frac{1}{4}m^2h\p_\n \e^\n+\half m\p_\m \e^\m \p_\n B^\n \nn\\
&&+\half m\p_\m \e^\m \p^2 \e+\half m^2\p_\m \e^\m \p_\n \e^\n-\frac{d}{8} m^2\e \p_\m B^\m \nn\\
&&-\frac{d}{8}m^2\e \p^2 \e-\frac{d}{8}m^3\e \p_\m \e^\m, \\
\frac{3}{2}m\eta \p_\m B^\m &\to& \frac{3}{2}m\eta \p^2 \e+\frac{3}{2}m^2\eta \p_\m \e^\m
-\frac{3}{2}m\p_\m B^\m \p_\n \e^\n \nn\\
&&-\frac{3}{2}m\p_\m \e^\m \p^2 \e-\frac{3}{2}m^2\p_\m \e^\m \p_\n \e^\n+\frac{9}{4}m^2\e\p_\m B^\m \nn\\
&&+\frac{9}{4}m^2 \e \p^2 \e+\frac{9}{4}m^3 \e \p_\m \e^\m
\eeq
\end{subequations}
It may be useful to rearrange the variations of $\mathcal{L}_{SZ}$ under the local gauge transformation in terms of powers of $m$ to explicitly show that they cancel out to the boundary terms.
\begin{itemize}
\item
Terms containing $h_{\m\n}$:
	\begin{subequations}
	\begin{itemize}
	\item Order of $m^0$
	\beq
	h_{\m \n}\p^2 \p^\m \e^\n-h_{\m \n}\p^\n \p^2 \e^\m+h_{\m \n}\p^\m \p^\n \p_\l \e^\l
	-h_{\m \n}\p^\m \p^\n \p_\l \e^\l=0.
	\eeq
	\item Order of $m^1$
	\beq
	-\frac{1}{4}mh\p^2 \e+\frac{1}{4}mh\p^2 \e+\half mh_{\m \n}\p^\m \p^\n \e-\frac{3}{2}mh_{\m \n}\p^\m \p^\n \e
	+mh_{\m \n}\p^\m \p^\n \e=0.
	\eeq
	\item Order of $m^2$
	\beq
	-m^2h_{\m \n}\p^\m \e^\n+m^2h_{\m \n}\p^\n \e^\m+\frac{1}{8}m^2h\p_\n \e^\n-\frac{3}{8}m^2h\p_\n \e^\n
	+\frac{1}{4}m^2h\p_\n \e^\n=0.
	\eeq 
	\item Order of $m^3$
	\beq
	\frac{1}{4}m^3h\e-\frac{d}{32}m^3h\e+\frac{9}{16}m^3h\e=\biggl(\frac{26-d}{32}\biggr)m^3h\e=0.
	\eeq
	\end{itemize}
	\end{subequations}

\item 
Terms containing $B_\m$:
	\begin{subequations}
	\begin{itemize}
	\item Order of $m^0$
	\beq
	B_\m \p^2\p^\m \e-B^\m \p_\m \p^2 \e=0.
	\eeq
	\item Order of $m^1$
	\beq
	&& mB_\m \p^2 \e^\m+m\p_\m \e_\n \p^\n B^\m+m\p_\n \e_\m \p^\n B^\m-m \p_\n \e^\n \p_\m B^\m
	-mB^\m \p_\m \p_\n \e^\n \nn\\
	&&+\half m\p_\m \e^\m \p_\n B^\n-\frac{3}{2}m\p_\m B^\m \p_\n \e^\n =0.
	\eeq
	\item Order of $m^2$
	\beq
	&&-\half m^2\e \p_\m B^\m+\frac{3}{2}m^2\e \p_\m B^\m-\frac{d}{8}m^2\e \p_\m B^\m+\frac{9}{4}m^2\e \p_\m B^\m = 			
	\biggl(\frac{26-d}{8}\biggr)m^2\e \p_\m B^\m=0.
	\eeq
	\end{itemize}
	\end{subequations}
	
\item 
Terms containing $\eta$:
	\begin{subequations}
	\begin{itemize}
	\item Order of $m^0$
	\beq
	2\eta \p^2 \p_\m \e^\m-\p_\m \e_\n \p^\m \p^\n \eta-\p_\n \e_\m \p^\m \p^\n \eta
	=0.
	\eeq
	\item Order of $m^1$
	\beq
	-3m\eta \p^2 \e+\half m\e \p^2\eta+m\eta \p^2 \e+\frac{3}{2}m\eta \p^2 \e
	=0.
	\eeq
	\item Order of $m^2$
	\beq
	-\frac{13}{4}m^2\p_\m \e^\m \eta+m^2\eta \p_\m \e^\m+\frac{3}{4}m^2\p_\m \e^\m \eta+\frac{3}{2}m^2\eta\p_\m \e^\m
	=0.
	\eeq
	\item Order of $m^3$
	\beq
	\frac{39}{8}m^3\eta \e-\frac{3d}{16}m^3\e \eta=\frac{3(26-d)}{16} m^3\eta \e=0.
	\eeq
	\end{itemize}
	\end{subequations}

\item
Terms not containing  fields:
	\begin{subequations}
	\begin{itemize}
	\item Order of $m^0$
	\beq
	&&\half \p_\m \e_\n \p^2 \p^\m \e^\n+\half \p_\m \e_\n \p^2 \p^\n \e^\m+\half \p_\m \e \p^2 \p^\m \e
	-\p_\m \e^\m \p^2 \p_\n \e^\n-\half \p_\m \e_\n \p^\m \p^\n \p_\l \e^\l \nn\\
	&&-\half \p_\n \e_\m \p^\n \p^2 \e^\m-\p_\m \e_\n \p^\n \p^2 \e^\m+\p_\m \e_\n \p^\m \p^\n \p_\l \e^\l
	+\p_\n \e_\m \p^\m \p^\n \p_\l \e^\l-\half \p^\m \e \p_\m \p^2 \e \nn\\
	&& 
	=0.
	\eeq
	\item Order of $m^1$
	\beq
	&&-\half m \p_\m \e^\m \p^2 \e+m\p_\m \e \p^2 \e^\m+3m\p_\m \e^\m \p^2 \e+m\e \p^2 \p_\m \e^\m
	-\frac{3}{2}m \p_\m \e_\n \p^\m \p^\n \e \nn\\
	&&-\frac{3}{2}m \p_\n \e_\m \p^\m \p^\n \e-\half m \e \p^2 \p_\m \e^\m+m \p_\m \e_\n \p^\m \p^\n \e
	+m\p_\m \e_\n \p^\m \p^\n \e-m\p_\n \e^\n \p^2 \e \nn\\
	&&-m\e^\m \p_\m \p^2 \e+\half m \p_\m \e^\m \p^2 \e-\frac{3}{2}m \p_\m \e^\m \p^2 \e 
	=0.
	\eeq
	\item Order of $m^2$
	\beq
	&&\frac{d}{16}m^2 \e \p^2\e-\half m^2 \p_\m \e_\n \p^\m \e^\n-\half m^2\p_\m \e_\n \p^\n \e^\m
	+\half m^2 \e_\m \p^2 \e^\m-\frac{9}{4}m^2 \e \p^2 \e \nn\\
	&&+\frac{13}{8}m^2\p_\m \e^\m \p_\n \e^\n-\frac{1}{8}m^2 \e \p^2 \e+\frac{3}{4}m^2 \e \p^2 \e+m^2\p_\n \e_\m \p^\n \e^\m
	-\half m^2 \e \p^2 \e \nn\\
	&&-m^2\p_\n \e^\n \p_\m \e^\m+\frac{3}{2}m^2 \e \p^2 \e+\frac{1}{8}m^2 \p_\m \e^\m \p_\n \e^\n
	-\half m^2 \e^\m \p_\m \p_\n \e^\n-\frac{3}{4}m^2 \p_\m \e^\m \p_\n \e^\n \nn\\
	&&+\half m^2\p_\m \e^\m \p_\n \e^\n-\frac{d}{8}m^2\e \p^2 \e-\frac{3}{2}m^2\p_\m \e^\m \p_\n \e^\n
	+\frac{9}{4}m^2 \e\p^2 \e \nn\\
	&=& \biggl(\frac{26-d}{16}\biggr)m^2\e \p^2 \e
	=0.
	\eeq
	\item Order of $m^3$
	\beq
	\biggl(\frac{41}{4}-\frac{41}{4}\biggr)\p_\m \e^\m \e=0.
	\eeq
	\item Order of $m^4$
	\beq
	\frac{(d-26)(d-18)}{128}m^4\e^2=0.
	\eeq
	
	\end{itemize}
	\end{subequations}
	
\end{itemize}
We find that the cancellations are highly non-trivial and that some terms cancel out only if the space-time dimension $d$ is equal to the critical value $d_{\rm critical} =26$. We recall here that this local gauge symmetry originates in the nilpotency of the BRST operator, which is applicable only in the critical dimension condition. It is interesting to note that even for the classical considerations, the local gauge invariance is valid only when $d=26$.

\end{document}